

\newif\iffigs\figsfalse
  \figstrue

\newif\ifamsf\amsffalse

\input harvmac
\iffigs
  \input epsf
\else
  \message{No figures will be included. See TeX file for more
information.}
\fi
\ifx\answ\bigans\else
  \message{This does not look very nice in little mode.}
\fi
\Title{IASSNS-HEP-93/4}
{\vbox{\centerline{ Multiple Mirror Manifolds and Topology Change }
   \vskip2pt\centerline{in String Theory}}}

\centerline{Paul S. Aspinwall,%
\footnote{${}^\dagger$}{
School of Natural Sciences, Institute for Advanced Study,
Princeton, NJ \ 08540.
}
Brian R. Greene,%
\footnote{${}^\sharp$}{
Schools of Mathematics and
 Natural Sciences, Institute for Advanced Study,
Princeton, NJ \ 08540.
On leave from: F.R. Newman Laboratory of Nuclear Studies,
Cornell University, Ithaca, NY \ 14853.
}
and David R. Morrison%
\footnote{${}^*$}{
School of Mathematics, Institute for Advanced Study,
Princeton, NJ \ 08540.
On leave from:  Department of Mathematics, Duke University,
Box 90320, Durham, NC \ 27708.
}}

\vskip .3in

\noblackbox

\font\eightrm=cmr8 \font\eighti=cmmi8
\font\eightsy=cmsy8 \font\eightbf=cmbx8
\font\eightit=cmti8 \font\eightsl=cmsl8 \skewchar\eighti='177
\skewchar\eightsy='60

\def\eightpoint{\def\rm{\fam0\eightrm}
\textfont0=\eightrm \scriptfont0=\fiverm \scriptscriptfont0=\fiverm
\textfont1=\eighti \scriptfont1=\fivei \scriptscriptfont1=\fivei
\textfont2=\eightsy \scriptfont2=\fivesy \scriptscriptfont2=\fivesy
\textfont\itfam=\eighti
\def\it{\fam\itfam\eightit}\def\sl{\fam\slfam\eightsl}%
\textfont\bffam=\eightbf \def\bf{\fam\bffam\eightbf}\rm}

\font\bigrm=cmr10 scaled \magstephalf
\def\inbar{\,\vrule height1.5ex width.4pt depth0pt}
\font\cmss=cmss10 \font\cmsss=cmss8 at 8pt
\def\BZ{\relax\ifmmode\mathchoice
{\hbox{\cmss Z\kern-.4em Z}}{\hbox{\cmss Z\kern-.4em Z}}
{\lower.9pt\hbox{\cmsss Z\kern-.36em Z}}
{\lower1.2pt\hbox{\cmsss Z\kern-.36em Z}}\else{\cmss Z\kern-.4em Z}\fi}
\def\IC{\relax\hbox{$\inbar\kern-.3em{\rm C}$}}
\def\IP{\relax{\rm I\kern-.18em P}}
\def\IQ{\relax\hbox{$\inbar\kern-.3em{\rm Q}$}}
\def\IR{\relax{\rm I\kern-.18em R}}
\ifamsf
  \font\bbl=msbm10
  \def\BZ{\hbox{\bbl Z}}
  \def\IC{\hbox{\bbl C}}
  \def\IP{\hbox{\bbl P}}
  \def\IQ{\hbox{\bbl Q}}
  \def\IR{\hbox{\bbl R}}
\fi

\def\Tr#1{\hbox{{\bigrm Tr}\kern-1.05em \lower2.1ex \hbox{$\scriptstyle#1$}}\,}

\def\ex#1{\hbox{$\> e^{#1}\>$}}

\def\WCP#1#2{\hbox{$\hbox{W\IC \IP}^{#1}_{#2}$}}

\def\CP#1{\hbox{$\hbox{\IC \IP}^{#1}$}}

\def\tilde{\widetilde}

\def\X{X}
\def\Y{\widehat{Y}}
\def\M{\widehat{X}}
\def\W{Y}

  We use mirror symmetry to establish the first concrete arena of spacetime
topology change in string theory. In particular, we establish that the
{\it quantum theories\/} based on certain nonlinear sigma models with
topologically distinct target spaces can be smoothly connected even though
classically a physical singularity would be encountered.  We accomplish this
by rephrasing the description of these nonlinear sigma models in terms of their
mirror manifold partners---a description in which the full quantum theory can
be described exactly using lowest order geometrical methods.
   We establish that, for the known class of mirror manifolds, the moduli space
of the corresponding conformal field theory requires not just two but
{\it numerous\/}  topologically distinct Calabi-Yau manifolds for its geometric
interpretation.  A {\it single\/} family of continuously connected conformal
theories thereby probes a host of topologically distinct geometrical spaces
giving rise to {\it multiple mirror manifolds}.

\Date{1/93}

\newsec{Introduction}

\nref\rGP{B.R. Greene and M.R. Plesser, Nucl. Phys. {\bf B338} (1990) 15.}%
\nref\rALR{P.S. Aspinwall, C.A. L\"utken, and G.G. Ross, Phys. Lett. {\bf 241B}
(1990) 373.}%
\nref\rAL{P.S. Aspinwall and C.A. L\"utken,
Nucl. Phys. {\bf B355} (1991) 482.}%
\nref\rnewW{E. Witten, ``Phases of $N=2$ theories in two dimensions'',
preprint IASSNS-HEP-93/3.}
\nref\rW{E. Witten, Nucl. Phys. {\bf B268} (1985) 79;
in {\it Essays on Mirror Manifolds}, (S.-T. Yau, ed.),
International Press Co., 1992, p. 120.}%
\nref\rDG{J. Distler and B. Greene,  Nucl. Phys.  {\bf B309} (1988) 295.}%
\nref\rAspin{P.S. Aspinwall, Comm. Math. Phys. {\bf 128} (1990) 593.}%

A venerable idea in field theory and in string theory is the possibility that
there might be physical processes whose description would involve a change in
the topology of the universe. Another long held
and partially realized hope is that string theory ameliorates the
disastrous physical consequences of spacetime singularities.
It is the purpose of the present letter to
show that each of these can be simultaneously realized in string theory
giving rise to {\it topology changing processes whose physical description is
completely smooth}.
We show  this in the
context of string theory as described by
conformally invariant  nonlinear sigma models with Calabi-Yau target spaces.
Our basic approach relies heavily on properties of mirror manifolds initially
emphasized in \refs{\rGP,\rALR,\rAL}.
An alternative approach, which gives a local description of the topology
changing process, has been simultaneously
pursued by Witten \rnewW\ using methods of
two-dimensional supersymmetric gauge theory.
The essential idea
in both approaches
is that a singular classical
operation such as topology change need not necessarily be singular when
quantum mechanical effects
 are taken into account. Investigating this, however,
can prove difficult as one typically needs to understand the {\it exact\/}
quantum mechanical description of a rather nontrivial physical situation
involving topology change.

Our approach to circumventing this obstacle involves two crucial properties of
such conformal theories. First, it has been known
for some time from the work of \refs{\rW,\rDG} that the
quantum mechanical  properties  of one sector of these conformal theories
(the ``complex structure sector'')  can
be {\it exactly\/} described using classical geometrical methods. Second,
mirror symmetry allows one to reinterpret such an exact quantum description
to exactly solve the complementary sector (the ``K\"ahler sector'') of the
conformal theory on the mirror manifold (for those cases in which the mirror
manifold is known to exist). Now, there is a natural
topology changing process%
\foot{Mirror symmetry itself associates two possible manifolds to a given
conformal field theory, but does so with an interchange of Hodge numbers.
The topology changing processes of interest here involve
no such change of Hodge numbers---with fixed Hodge numbers, the topology
of the background is changing as the parameters vary.}
in the K\"ahler sector  supplied by algebraic geometry in which two
topologically distinct manifolds are linked by a common singular intermediate
space. By following the above outline, we can exactly describe the
quantum mechanical version of this procedure as a particular operation
on the complex structure of the mirror manifold.
An examination of this quantum
theory reveals that {\it no singularity is encountered\/}%
\foot{Provided that one follows a sufficiently general path.}
when passing through
the topology change on the original space---in fact, in the mirror
geometrical description no topology change occurs at all. In this way we
see that the underlying conformal theory, which governs the physics,
quantum mechanically ``smooths'' out the singular classical behavior
associated
with topology change.
In \refs{\rAspin,\rAL} it was realized that string theory probably provides
some
smoothing of these singularities. Here we are able to establish precisely what
happens.
Furthermore,  our results show that rather than being
geometrically interpretable in terms of
two topological types of Calabi-Yau spaces, the underlying single family of
conformal theories requires numerous distinct topological types for its
geometrical description. One might say that
these {\it multiple mirror manifolds\/} give rise
to a kind of {\it catoptric\/} phenomenon in mirror symmetry.

\nref\rCGH{P. Candelas, P.S. Green, and T. H\"ubsch, Nucl. Phys. {\bf B330}
(1990) 49.}%

It is worth emphasizing that the types of topology change considered
here differ significantly from the ``conifold transitions'' studied
in \rCGH.  Whereas our topology changing processes are physically
smooth, those of \rCGH\ encounter singularities which make them
difficult (perhaps impossible) to interpret physically.

\nref\rAGMP{P.S. Aspinwall, B.R. Greene, D.R. Morrison, and M.R. Plesser,
in preparation.}%

The body of the present letter is concerned with presenting strong
evidence in favor of the picture outlined above. We do this in the context
of an extensive investigation of a particularly tractable example which
illustrates the phenomena we have described.
In section 2 we give a brief review of the aspects of mirror symmetry
relevant to our discussion and
more clearly describe the specific question under study. In section 3
we present
the example we  study and give
the explicit calculation which verifies
the picture we are presenting. In section 4
we briefly give our conclusions.
A more detailed description of the present work along with an
investigation of a number of important related issues will be given
in \rAGMP.

\newsec{Mirror Symmetry and Flops}

\nref\rDixon{L.J. Dixon, in {\it Superstrings, Unified Theories, and
Cosmology 1987}, (G. Furlan et. al., eds.), World Scientific, 1988, p. 67.}%
\nref\rLVW{W. Lerche, C. Vafa, and N.P. Warner, Nucl. Phys. {\bf B324} (1989)
427.}%
\nref\rCLS{P. Candelas, M. Lynker, and R. Schimmrigk, Nucl. Phys. {\bf B341}
(1990) 383.}%
\nref\rORB{L. Dixon, J.A. Harvey, C. Vafa, and E. Witten, Nucl. Phys.
{\bf B261} (1985) 678; Nucl. Phys. {\bf B274} (1986) 285.}

Mirror symmetry describes a situation in which a single moduli (sub-)space
of conformal field theories, all related to one another by truly marginal
deformations, has {\it two\/} geometrical interpretations.
These interpretations are in terms of
two families of topologically distinct  Calabi-Yau manifolds,
with members of a given family related to one another by deformations of
the complex and/or K\"ahler structure. Mirror symmetry was conjectured
based upon naturality arguments in \refs{\rDixon,\rLVW},
was strongly suggested by the computer studies of \rCLS,
and was established to
exist in certain cases
by direct construction in \rGP. The construction of \rGP, which will
play a central role in our study, states that if $\X$ is an
$n$-dimensional Calabi-Yau
hypersurface of Fermat type in weighted projective space $\WCP{}{}$ and if $G$
is
the maximal subgroup of diagonal scaling symmetries on the homogeneous
$\WCP{}{}$ coordinates that preserves $\X$ and leaves the holomorphic $(n,0)$
form
on $\X$ invariant, then $\X$ and $\X/G$ constitute a mirror pair.
An important point to notice is that, typically, both $\X$
and $\X/G$ are singular spaces. The reason for this is twofold. First, the
ambient weighted projective space has singularities and if $\X$ passes
through them, it will be singular as well. Second, even if $\X$ were smooth,
$G$ acts with fixed points (unless $n = 1$) and hence $\X/G$ has orbifold
singularities. Although treacherous from the point
particle viewpoint, work over the last few years
has established \rORB\ that string theory is perfectly well defined
on such singular spaces.

\nref\rKollar{J. Koll\'ar, Nagoya Math. J. {\bf 113} (1989) 15.}%

Our interest, however, is not solely on the isolated
point in conformal field theory moduli space associated with $\X$ and $\X/G$,
but rather with the whole moduli space to which this point belongs.
The study of this space
raises the following interesting issue. It is well
known  that the geometrical singularities encountered in these
examples can be repaired (in a manner that preserves the Calabi-Yau condition
of vanishing first Chern class) but that there is not a {\it unique\/} way
of doing so.
That is, the blowing-up procedure which is used to desingularize one of
these spaces (say, $X$)
gives rise to {\it numerous\/} possible smooth models $\M$ which
are topologically distinct resolutions of the original singular space.
Each of these possible resolutions differs from the others by a series of
``flops'' of rational curves \rKollar: a procedure in which a rational curve on
the
smooth variety is blown down to a point (yielding a singular space) and then
blown up (to yield a smooth space once again) in a manner which changes the
topology of the resulting manifold.

The natural question to ask
is whether one, or some subset,
of these resolutions (which are
on equal footing mathematically) is somehow picked out physically by the string
theory or whether all of the possible topologically distinct
resolutions are realized by physical
models. Although at first sight this question might seem  like a technicality
it is in fact of deep significance to our understanding of nonlinear sigma
models, to the crucial issue of how the string copes with singular spaces,
and to the question of topology change.

To focus on this question more directly, notice that
the operation of flopping can be thought of as a particular
motion in the K\"ahler moduli space of the variety and its corresponding
conformal field theory.
More specifically,
a nonlinear sigma model with a smooth Calabi-Yau
target space $\M$ gives rise to an $N = 2$ superconformal field theory.
This conformal theory has a number of truly marginal operators which are
geometrically interpretable in terms of deformations of the complex or
K\"ahler structure of the associated Calabi-Yau manifold. Since the K\"ahler
moduli determine the metric on the manifold, there are criteria which these
moduli must meet in order that the corresponding classical metric be
nonsingular. Writing the conformal field theory K\"ahler modulus field $K$
in the standard form $K = B + iJ$ with $J$ being in the cohomology class of
the K\"ahler form on $\M$,
then these requirements are%
\foot{We will focus on the threefold case for the
most part even though our discussion is largely independent of the
dimension.}

\eqn\eJJJ{ \int_{\M} J \wedge J \wedge J > 0, \qquad \int_S J \wedge J   > 0,
\qquad \int_C J > 0}
where $S$ and $C$ are homologically nontrivial hypersurfaces and curves
in $\M$, respectively. These conditions restrict $J$ to a cone-like region
which is commonly known as the {\it K\"ahler cone}.

If $C_0$ is a curve which can be flopped, then $C_0$ determines a wall
of the K\"ahler cone, on which $\int_{C_0}J=0$ but the other constraints
of \eJJJ\ continue to be satisfied.  As the imaginary part $J$ of
the K\"ahler moduli field approaches that wall, the volume of the
curve $C_0$ shrinks towards 0.  Along the wall itself, the natural
geometrical model will be the singular one in which the curve $C_0$
has been blown down to a point.  If $J$ actually crosses the wall,
the volume of $C_0$ would
appear to have become negative; however, the effect of the flopping operation
is to restore the positivity of the volume of the curve, but on a
different topological model of the space.  We can think of this
flopping operation of classical algebraic
geometry
as providing
a means of traversing a wall of the K\"ahler
cone by passing
through a {\it singular\/} space and then on to a different smooth
topological model.  This suggests that the K\"ahler moduli spaces
of the distinct resolutions should be attached along walls to form
a larger moduli space.

To form a better picture of what such an attachment would look like,
we pass to a more natural set of coordinates on the K\"ahler moduli space.
Notice that the
underlying conformal field theory is invariant under shifts in the
antisymmetric tensor field $B$ by elements of integral cohomology.
The moduli space should therefore only have a single point for the class of
theories differing by integral shifts.  This observation can be implemented
by exponentiating the na\"\i ve coordinates, and
considering the natural coordinates on the moduli space to be
$w_k =  e^{2 \pi i (B_k + iJ_k)}$,
where $B_k$ and $J_k$ are the components of the two-forms $B$ and $J$
relative to an integral basis of $H^2(\M,\BZ)$.
(Thus $J=-{1\over2\pi}(\log|w_1|,\dots,\log|w_r|)$, which we abbreviate
as $J=-{1\over2\pi}\log|\vec{w}|$.)
The restriction that $J$ lie
in the K\"ahler cone is mapped into the statement
that $\vec{w}$ lie in a bounded domain (an intersection of polydisks).
Our analysis of how a wall in the K\"ahler cone can be traversed by means
of a flop leads immediately to the idea
of attaching the closures of
these bounded domains along a common boundary (a ``wall'').
Attaching these regions for all possible resolutions leads to a
``partially enlarged moduli space''
which we schematically depict in figure 1.
(In \rAGMP\ we will discuss the ``fully enlarged moduli space'' which
in addition to the above, contains numerous other regions,
not obtained by flopping curves on smooth manifolds.  Similar
additional regions also arise from the point of view of Witten's
analysis \rnewW\
and almost certainly describe the same physical systems.)
\iffigs
$$\vbox{\centerline{\epsfxsize=4cm\epsfbox{mult-f1.ps}}
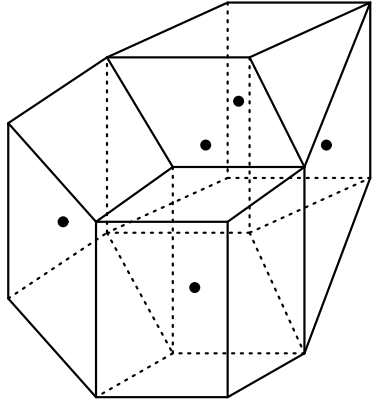
\centerline{Figure 1. The partially enlarged moduli space.}}$$
\fi

The question we seek to answer regards
the physical significance of the regions in the partially enlarged
moduli space.
{\it A priori\/} one can imagine two possible answers. First, it may be that
 only one
(or some proper subset)
of these possible regions has a physical interpretation. Mathematically this
seems unnatural as there is no fundamental way to select one resolution
(or a proper subset) from
all of the possibilities---but it is conceivable that the string has its own
inherent way of resolving singularities.
If this possibility were to be correct, it would mean that the regions
determined by a single
(complexified) K\"ahler cone would correspond, under mirror symmetry, to
the full complex structure moduli space of the mirror.
In particular, the geometric flopping operation described above would
not be realized in conformal field theory.
Alternatively, it may be that  {\it all\/} of the possible regions are realized
by
physical models.
We shall refer to this second possibility as
giving rise to {\it multiple mirror
manifolds\/}---reflecting the fact that it implies that a single
conformal field theory moduli space requires not two but numerous
topological types for its geometric interpretation.
This possibility is not immediately convincing since
the walls dividing the moduli space into regions as in figure 1 are
not directly visible
in the corresponding conformal field theory moduli space.
Equivalently, invoking mirror symmetry, such a
structure of walls and regions
is not visible in the complex structure moduli space of the mirror manifold%
\foot{Although not of direct relevance to our present study, for completeness
we note a third possibility: maybe possibility
two is correct and all possible regions are physically
realized but there  is an underlying
symmetry which identifies---at the level of conformal field theory---a
point in one region with a point (or points) in each of the other regions
thus showing that the true physical moduli space has a fundamental domain which
resides in one region.}.

Fortunately, by invoking mirror symmetry, there is a concrete calculation
which can be used to adjudicate between these possibilities---and, in
particular, give extremely persuasive evidence that the multiple mirror
manifold option is correct.
We now describe this calculation.

\nref\rStromWitten{A. Strominger and E. Witten, Comm. Math. Phys. {\bf 101}
(1985) 341.}
\nref\rStrom{A. Strominger, Phys. Rev. Lett. {\bf 55} (1985) 2547.}%
\nref\rDSWW{M. Dine, N. Seiberg, X.-G. Wen, and E. Witten, Nucl. Phys.
{\bf B278} (1987) 769; Nucl. Phys. {\bf B289} (1987) 319.}%
\nref\rCDGP{P. Candelas, X.C. de la Ossa, P.S. Green, and L. Parkes,
Phys. Lett. {\bf 258B} (1991) 118; Nucl. Phys. {\bf B359} (1991) 21.}%
\nref\rAM{P.S. Aspinwall and D.R. Morrison, ``Topological field theory
and rational curves'', Comm. Math. Phys., in press.}%

Let $\M$ and $\Y$ be a mirror pair of Calabi-Yau manifolds. Because
they are a mirror pair, a striking and extremely useful equality between
the Yukawa couplings amongst the $(1,1)$ forms on $\M$ and the $(2,1)$ forms
on $\W$ (and vice versa) is satisfied. This equality demands that \rGP:
\eqnn\eEQUAL
$$\displaylines{
\int_{{\Y}} \omega^{abc}
\tilde b^{(i)}_a \wedge \tilde b^{(j)}_b \wedge \tilde b^{(k)}_c
\wedge \omega  = \hfill\eEQUAL \cr
\int_{{\M}} b^{(i)} \wedge b^{(j)} \wedge b^{(k)} +
\sum_m \sum_{\{\Gamma \}}\ex{\int_{\Gamma }u_m^*(K)}
 ( \int_{\Gamma } u^*(b^{(i)} )) ( \int_{\Gamma } u^*(b^{(j)} ))
( \int_{\Gamma } u^*(b^{(k)} ))\> .}$$
where on the left hand side (as derived in \rStromWitten)
the $\tilde b^{(i)}_a$ are $(2,1)$ forms (expressed as elements
of $H^1(\Y,T)$ with their subscripts being tangent space indices),
$\omega$ is the holomorphic three form and on the right hand side
(as derived in
\refs{\rStrom,\rDSWW,\rCDGP,\rAM})
the $b^{(i)}$ are $(1,1)$ forms on $\M$,
$\{\Gamma \}$ is the set of rational curves on $\M$,
$u: \IC\IP^1 \rightarrow \Gamma $ is a holomorphic map to a rational curve,
$\pi_m$ is an $m$-fold cover $\IC\IP^1 \rightarrow \IC\IP^1$ and
$u_m = u \circ \pi_m$.
It is important to remark that
this equation is equally valid if $\Y$ is replaced by a singular space
$\W$, as the integral on
the left hand side of \eEQUAL\
is insensitive to singularities of $\W$.

\nref\rM{D.R. Morrison, in {\it Essays on Mirror Manifolds}, (S.-T. Yau, ed.),
International Press Co., 1992, p. 241.}%

Notice the interesting fact that at ``large radius''%
\foot{``Large radius'' here refers to limit points in the
K\"ahler moduli space in which all instanton corrections
are suppressed.}
the right hand side
of \eEQUAL\ reduces to the topological intersection form on $\M$ and
hence mirror symmetry (in this particular limit) equates a topological
invariant of $\M$ to a quasitopological invariant (i.e., one depending on the
complex structure) of $\W$ \rGP.
If the multiple mirror manifold option is correct,
and all regions of figure 1 are physically realized,
 the following must hold:
Each of the distinct intersection forms,
which represents the large radius limit of the $(1,1)$ Yukawa couplings
on each of the topologically distinct resolutions of $\X$, must be
equal to the $(2,1)$ Yukawa couplings on $\W$ for suitable corresponding
``large complex structure'' limits. We schematically illustrate these limits
by means of the marked points in the interiors of the regions
in figure 1.
Now, actually invoking this equality
requires understanding the precise complex structure limit that corresponds,
under mirror symmetry to the particular large K\"ahler structure limit
being taken%
\foot{In \rALR\ the first explicit realization and verification of
\eEQUAL\ (for a mirror pair based on the quintic in $\CP4$)
in a large radius limit was carried out by means of a well motivated
guess for the corresponding large complex structure limit point.
 In
\rCDGP\ equation \eEQUAL\ was combined with an explicit determination of the
{\it mirror map\/} (the map between the K\"ahler moduli space of $\X$ and the
complex structure moduli space of $\X/G$ for $\X$ being the Fermat
quintic in $\CP4$)
to determine the number of rational curves of arbitrary degree on
(deformations of) $\X$. This
calculation was subsequently described mathematically and extended to
a number of other examples in \rM.}.
In the following section we shall find these limit points in  the
complex structure moduli space
 and  verify that each corresponds to the mirror
of a {\it topologically  distinct\/} large radius Calabi-Yau manifold.

Furthermore, we can follow paths in the complex structure moduli space
which connect these complex structure limit points in a manner that
encounters no physical singularity.  The reason for this is that
in the complex structure description, singularities arise only if the
variety is not transverse (equivalently, if the Landau-Ginzburg superpotential
does not have an isolated critical point at the origin of field space).
This occurs in a {\it complex\/} codimension one subspace of the
moduli space and hence can always be avoided
by a judicious choice of path.
Part of the difference in the
K\"ahler vs.\ complex structure description is due to the fact that classical
singularities in the former are solely dependent on the imaginary part of
the K\"ahler modulus and hence occur in {\it real\/} codimension one giving
rise
to the wall like structure we have  discussed above.
Thus whereas the classical (K\"ahler) description involves topology change
through a singular space, the quantum description is perfectly smooth.

\newsec{The Example and The Calculation}

\nref\rRoan{S.-S. Roan, Int. Jour. of Math. {\bf 2} (1991) 439.}%
\nref\rBatyrev{V.V. Batyrev, ``Dual polyhedra and mirror symmetry
for Calabi-Yau hypersurfaces in toric varieties'', Essen preprint,
November 18, 1992.}
\nref\rToric{T. Oda, {\it Convex Bodies and Algebraic Geometry},
Springer-Verlag, 1988.}

We choose $\X$ to be given by the vanishing locus of
\eqn\eEQN{z_0^3 + z_1^3 + z_2^6 + z_3^9 + z_4^{18}} in
\WCP4{6,6,3,2,1}.
The construction of \rGP\ gives us a mirror partner of $\X$ of the form
$\X/G$ where $G$ is the discrete group $\BZ_3 \times \BZ_3
\times \BZ_3$ acting by
\eqn\eOrb{(z_0,z_1,z_2,z_3,z_4) \rightarrow
 (\alpha \beta \gamma\, z_0, \alpha^{-1} z_1,\beta^{-1}z_2, \gamma^{-1}z_3,
z_4)}
with $\alpha$, $\beta$, and $\gamma$ cube roots of unity.
The nontrivial Hodge numbers of resolutions $\M$ of
this mirror pair are $h^{1,1}_{\M} = 7$ and
$h^{2,1}_{\M} = 79$.
The calculation outlined in the previous section will make use of a five
dimensional subspace of the
$(1,1)$ cohomology of $\M$ and
a corresponding five dimensional subspace of the $(2,1)$ cohomology of $\X/G$.
We now describe each of these in turn. We will use the language of toric
varieties in our presentation as this is the appropriate mathematical
forum for describing mirror symmetry in these examples
\refs{\rRoan,\rBatyrev}. It is also the correct framework for discussing
desingularizations and the structure of K\"ahler moduli space.
The reader unfamiliar with toric geometry should consult
\rToric\ and might, on a first reading, wish to skip directly
to subsection ``Predictions and their verification''.

\subsec{The K\"ahler sector of $\M$}

As discussed in section 2, $\X$ is singular because it intersects
the singularities of the ambient weighted projective space. It is easily
verified that the singularities of $\X$ consist the two curves
$z_3 = z_4 = 0$ and $z_2 = z_4 = 0$ which themselves intersect in three points.
These curves are $\BZ_2$ and $\BZ_3$ fixed point sets, respectively, and
the points of intersection are $\BZ_6$ fixed point sets. To resolve these
singularities, as mentioned, it is most efficient to use the methods of
toric geometry \refs{\rToric,\rAspin}.
 In this setting, we resolve the singularities
on the ambient weighted space rather than working directly on $\X$.

\nref\rnewRoan{S.-S. Roan, ``Topological properties of Calabi-Yau mirror
manifolds'', Max-Planck-Institut f\"ur Mathematik preprint.}

To do so, let ${\bf N}$ denote the the algebraic group homomorphisms
 from $\IC^*$ to the four dimensional algebraic torus ${\cal T}$.
$ \bf N$ is isomorphic to the lattice $\BZ^{4}$. Let $\bf M$ be the dual
lattice which represents the group characters of ${\cal T}$.
A toric variety is specified by giving a fan decomposition of
all or part of $\bf N\otimes\IR$. A family of hypersurfaces $\X_t$ in
a toric variety (such as the Calabi-Yau hypersurfaces in weighted projective
space) can be specified by choosing a polyhedron $\Delta$ in $\bf M$
which determines the collection of monomials appearing in the equations
of the hypersurfaces.  (To describe a particular hypersurface $\X$,
specific coefficients for those monomials must be chosen to
form the equation.)
The corresponding dual polyhedron $\Delta^*$ in $\bf N$   contains
information relevant to the K\"ahler structure of the hypersurface.
A combinatorial version of
mirror symmetry
\rBatyrev\
states that the family of mirrors of a Calabi-Yau
hypersurface $\X$
(with varying K\"ahler structure)
should be
obtained by building the family of Calabi-Yau hypersurfaces $\W_s$
which corresponds
to the data given by interchanging the roles of $\bf N$ and $\bf M$%
\foot{
The equality of Hodge numbers of resolutions of
$\X$ and $\W$ has been verified by
Roan \refs{\rRoan,\rnewRoan} and Batyrev \rBatyrev, lending strong support
to the conjecture that $\X$ and $\W$ in fact form a mirror pair.  However,
the power of mirror symmetry derives from there being
an underlying common conformal field theory for the pair of
mirror manifolds---as yet these mathematical descriptions
fall far short of supplying this
crucial link. Our study here, of course, relies on the conformal field theory
link established in \rGP.}.
When $\X$ is a Fermat hypersurface as in our example, one such $\W$
is given by $\X/G$.

The fan consisting of cones over faces of $\Delta^*$ determines a fundamental
toric variety containing the hypersurfaces $\X_t$, which in our case coincides
with the weighted projective space \WCP4{6,6,3,2,1}.
Following the discussion in \rBatyrev,
the vertices of $\Delta^*$ are straightforwardly determined to
be
\eqn\eVertDeltastar{
                     \nu_1 = (1,0,0,0),
                     \nu_2 = (0,1,0,0),
                     \nu_3 = (0,0,1,0),
                     \nu_4 = (0,0,0,1),
                     \nu_5 = (-6,-6,-3,-2) }
and the origin $n_0=(0,0,0,0)$ is the unique lattice point in the interior
of $\Delta^*$.
We let $T_i$ denote the cone with tip at the origin and having a polyhedral
base consisting of all points in \eVertDeltastar\ except $\nu_i$.
The 4-dimensional cones in our fan are exactly $T_1$,\dots,$T_5$, and for
the corresponding space to be smooth the base of each must have unit volume.
Equivalently,  the base must not contain any other
lattice point of $\bf N$. Direct examination reveals this, as expected, not
to be the case as the lattice points
\eqn\ePointsDeltaStar{
                     n_1 = (-3,-3,-1,-1),
                     n_2 = (-2,-2,-1,0),
                     n_3 = (-4,-4,-2,-1),
                     n_4 = (-1,-1,0,0)  }
are contained in some of our bases.

To desingularize this toric variety, we need to subdivide our five cones
so that each of the $n_i$ is contained in the base of some tetrahedral cone,
i.e.\ so that each resulting (truncated) cone has unit volume.
This requires that we subdivide $T_1$ and $T_2$ into six cones,
$T_3$ into three cones and $T_4$ into two cones. Each such cone
subdivision is specified by a subdivision of the base. Notice, though, that
whereas there is a unique subdivision of cones $T_3$ and $T_4$, there are
{\it five\/} possible subdivisions of cones $T_1$ and $T_2$.  We show these
five possible resolutions of the singular variety in figure 2
along with
the curve flops which take us from one resolution to another (using the
notation $n_in_j$ to indicate which segment in the graph corresponds to
the curve which should be flopped). All of these
resolutions are K\"ahler.
\iffigs
$$\vbox{\centerline{\epsfxsize=9cm\epsfbox{mult-f2.ps}}
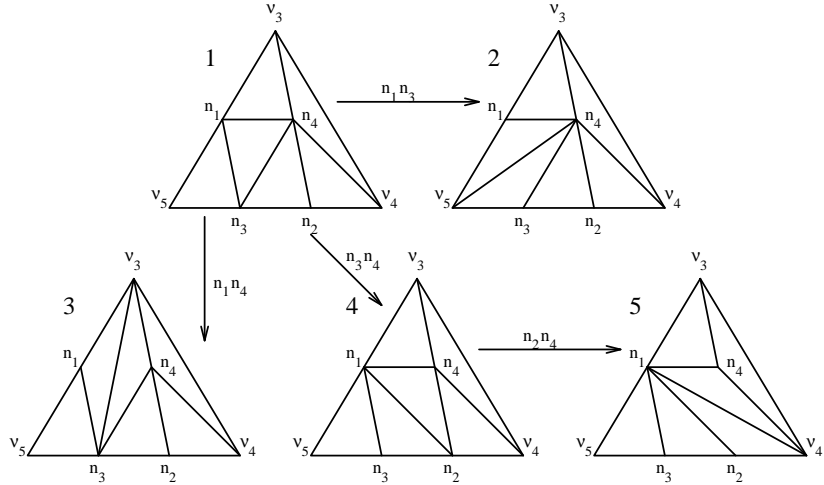
\centerline{Figure 2. The five possible resolutions.}}$$
\fi

The lattice points
$n_1$,\dots,$n_4$
give rays in ${\bf N} \otimes \IR$ which correspond to the
exceptional divisors $\widetilde{E}_1$,\dots,$\widetilde{E}_4$
on any resolution of the weighted projective space. The intersection%
\foot{$\widetilde{E}_4$ meets $\M$ in a divisor with three components,
corresponding to the three points of intersection of the two curves of
singularities; we take $E_4$ to be the {\it average\/} of those three
divisors on $\M$.}
of these exceptional divisors with $\M$ (the proper transform of $\X$
on the resolution) gives
four divisors $E_1$,\dots,$E_4$ which together with the hyperplane section
$H$ from the original weighted projective space span a five-dimensional
space of divisors, Poincar\'e dual to the five-dimensional subspace
of $H^{1,1}(\M)$ which
we shall study. The other two elements (recall $h^{1,1}_{\M} = 7$) are not
realizable in terms of toric divisors and can effectively be ignored for the
purposes of this letter.

The intersection numbers for each of these five resolutions can be
calculated from the corresponding toric diagram.  In table 1 we list
the particular nontrivial
 invariant ratios of intersection numbers%
\foot{These ratios are independent of the (uncomputed)
 Zamolodchikov metric for these theories, if this metric is diagonal
in the large radius/complex structure limit we consider. This can be
strongly motivated by a consideration of discrete symmetries as in \rALR\
or, more precisely, by an examination of the Weil-Petersson metric. In fact,
it appears that such an examination reveals a particular form for this metric
which allows one to go further than checking invariant ratios as we shall
discuss in \rAGMP.},
for each resolution, which
are relevant to our present calculation.
For certain resolutions some of these ratios are indeterminate and this is
represented in the table by $0/0$. The data in the table, ignoring the
indeterminate entries, are sufficient to distinguish between the five
resolutions.%

\def\tablerule{\noalign{\hrule}}
\def\ilspace{\omit&height2pt&&&&&&&&&&&&\cr}
$$\vbox{\tabskip=0pt \offinterlineskip
\halign{\strut#&
\vrule#&\hfil\quad$#$\quad\hfil&\vrule#&\hfil\quad$#$\quad\hfil&
\vrule#&\hfil\quad$#$\quad\hfil&\vrule#&\hfil\quad$#$\quad\hfil&
\vrule#&\hfil\quad$#$\quad\hfil&\vrule#&\hfil\quad$#$\quad\hfil&
\vrule#\cr \tablerule\ilspace
&&\hbox{Resolution}&&1&&2&&3&&4&&5&\cr\ilspace\tablerule\ilspace
&&{(E_1^3)(E_4^3)\over(E_1^2E_4)(E_1E_4^2)}
&& -7 && 0/0 && 0/0 && \infty && 9 &\cr\ilspace
&&{(E_2^2E_4)(E_3^2E_4)\over(E_2E_3E_4)(E_2E_3E_4)}
&& 2 && 4 && 0 && 0/0 && 0/0 &\cr\ilspace
&&{(E_2E_3E_4)(HE_2^2)\over(E_2^2E_4)(HE_2E_3)}
&& 1 && 1 && 1 && 0 && 0/0 &\cr\ilspace
&&{(E_2E_3E_4)(HE_1^2)\over(E_1^2E_4)(HE_2E_3)}
&& 2 && 1 && \infty && 0/0 && 0 &\cr\ilspace
\tablerule
\noalign{\vskip2pt}
&\multispan{13}\hfil Table 1: Ratios of intersection numbers\hfil\cr
}}$$

\subsec{The complex structure sector of $\X/G$}

In order to study the complex structure moduli space of $X/G$, we need
not resolve singularities. (Although there are numerous topologically
distinct resolutions of $X/G$, the differences among them only affect
the K\"ahler sector.)
Five of the seven $(2,1)$ forms on $\X/G$ have a monomial representation
(the other two arise in twisted sectors of the orbifold
theory).
These are the five of the original seventy-six monomial deformations on $\X$
(although $h^{2,1}_{\M} = 79$,
only $76$ have a monomial representation)
which are invariant under $G$ and are explicitly given in table 2.
In this table we also give a correspondence between these five $(2,1)$ forms
on $\X/G$ and the five $(1,1)$ forms on its mirror, $\M$. Physically, the
work of \rGP\ establishes that such a correspondence exists (each massless
mode in the single underlying conformal theory has two geometrical
interpretations thus establishing the correspondence). Mathematically, this
explicit correspondence, which we shall refer to as the ``monomial-divisor
mirror map'' is inspired by the work of \rRoan\ and \rBatyrev\ in which
the interpretation of the $\bf M$ lattice of $\X/G$ as the $\bf N$ lattice
of $\X$
and vice versa
(the toric description of mirror manifolds) gives rise to this
natural association of divisors on $\M$ to monomial deformations on $\X/G$.

\def\ilspace{\omit&height2pt&&&&&&\cr}
$$\vbox{\tabskip=0pt \offinterlineskip
\halign{\strut#&
\vrule#&\hfil\quad#\quad\hfil&\vrule#&\hfil\quad#\quad\hfil&
\vrule#&\hfil\quad#\quad\hfil&
\vrule#\cr \tablerule\ilspace
&&$(2,1)$-form&&Monomial && Divisor &\cr\ilspace
\tablerule\ilspace
&&$\varphi_0$ &&$ z_0z_1z_2z_3z_4$ &&$H$&\cr\ilspace
&&$\varphi_1$ &&$ z_2^3z_4^9$ &&$E_1$&\cr\ilspace
&&$\varphi_2$ &&$ z_3^6z_4^6$ &&$E_2$&\cr\ilspace
&&$\varphi_3$ &&$ z_3^3z_4^{12}$ &&$E_3$&\cr\ilspace
&&$\varphi_4$ &&$ z_2^3z_3^3z_4^3$ &&$E_4$&\cr\ilspace
\tablerule\noalign{\vskip2pt}
&\multispan{7}\hfil Table 2: Monomial-divisor mirror map\hfil\cr
}}$$

\nref\rCandelas{P.~Candelas, Nucl. Phys. {\bf B298} (1988) 458.}%

The general point in this five dimensional subspace of the complex
structure moduli space of $\X/G$ with coordinates $(a_0,\dots,a_4)$
corresponds to the complex structure
\eqn\eComplex{z_0^3 + z_1^3 + z_2^6 + z_3^9 + z_4^{18} + a_0 z_0z_1z_2z_3z_4 +
a_1 z_2^3z_4^9 + a_2 z_3^6z_4^6 + a_3 z_3^3z_4^{12} + a_4 z_2^3z_3^3z_4^3 = 0.}
The monomial-divisor mirror map in fact goes farther:  it asserts that
the mirror isomorphism between this moduli space and the enlarged K\"ahler
moduli space of $\M$ should be given by $w_i=1/a_i$ (up to constants),
at least asymptotically near the large radius limit points.

The three point functions amongst the $(2,1)$ forms on $\X/G$ with
equation given by \eComplex\ (for arbitrary
$(a_0,\dots,a_4)$) are in principle calculable using the method developed in
\rCandelas. In practice, deriving a closed form expression in terms of
the $a_i$ is a formidable calculational task which appears to be beyond present
computer capabilities. Luckily, for any chosen numerical values of the
$a_i$, say integers,
we can directly determine the corresponding three point functions exactly
using Maple.

Now, as discussed earlier, if each possible resolution is to have
a corresponding distinct physical model, for each
resolution we should be able to choose the
$a_i$ so that the corresponding $(2,1)$ couplings agree with the
$(1,1)$ couplings in table 1. More precisely, ``large complex structure''
corresponds \refs{\rALR,\rCDGP} to the $a_i$ getting large in absolute value,
and so to the $w_i=1/a_i$ staying small.
The distinction between different large complex structure limits derives
 from the various regions which
$-{1\over2\pi}\log|\vec{w}|={1\over2\pi}\log|\vec{a}|$
may inhabit---different
regions will lead to different limits.  What changes from region to
region is the {\it relative\/} rates of growth of the $a_i$.
In practice, we write $ a_i  = s^{r_i} $ and our goal is to find
five distinct ``direction''
vectors $\vec r = (r_1,r_2,r_3,r_4,r_5)$ which lie in the
regions corresponding to the five
resolutions of $\X$.

\subsec{The secondary polytope}

\nref\rOdaPark{T. Oda and H.S. Park, T\^ohoku Math. J. {\bf 43} (1991) 375.}%
\nref\rGKZ{I.M. Gel'fand, A.V. Zelevinskii, and M.M. Kapranov, Leningrad
Math. J. {\bf 2} (1991) 449.}%
\nref\rCornell{L.J. Billera, P. Filliman, and B. Sturmfels, Advances in Math.
{\bf 83} (1990) 155.}%
\nref\rnewBatyrev{V.V. Batyrev, ``Variations of the mixed Hodge structure
of affine hypersurfaces in algebraic tori'', Essen preprint, August 1992.}%

In section 2 we schematically described the K\"ahler moduli space
of the kinds of varieties we are discussing as consisting of a collection
of K\"ahler cones corresponding to varieties obtained from one another by
a series of flops. We have further indicated how, in the present setting,
each such region is the K\"ahler moduli space for one possible resolution
of the underlying singular space.
Oda and Park \rOdaPark\ have shown how to calculate the K\"ahler
cones for each resolution (associated to a triangulation of $\Delta^*$),
based on the ``secondary polytope'' construction of \rGKZ\ (see also
\rCornell)%
\foot{The possibility of using the secondary polytope construction to
find ``large complex structure'' limits was anticipated by
Batyrev \rnewBatyrev.}.
We will describe the algorithm in detail in \rAGMP; when applied to
the case at hand, we find that each K\"ahler cone is determined by
a (redundant) set of 90 linear inequalities.  Solving these inequalities,
we find that the cones are actually simplicial, so each can be described as
the set of all positive linear combinations of a collection of 5 generators.
These generators are given in table 3.  (The monomial-divisor mirror map
has determined a natural set of coordinates on the enlarged K\"ahler
moduli space, used in table 3a.)  We have illustrated the way in which
these cones fit together in figure 3.
\foot{All of our 5-dimensional cones share a common generator $v_1$,
so we have depicted the 4-dimensional cones generated by the remaining
edges.  We show these 4-dimensional cones by displaying their intersection
with the 3-sphere $S^3$ (which is itself flattened to $\IR^3$ in the
figure).  One can regard the polyhedral regions in the figure as the
bases of the cones, which must be connected to a vertex to
obtain cones.}

{\ninepoint
\def\ilspace{\omit&height2.5pt&&&&\cr}
$$\vbox{\tabskip=0pt \offinterlineskip
\halign{\strut#&
\vrule#&\hfil\quad$#$\quad\hfil&\vrule#&\hfil\quad$#$\quad\hfil&
\vrule#\cr \tablerule\ilspace
&&v_1&&(-{1\over3},0,0,0,0)&\cr\ilspace
&&v_2&&(-{7\over{18}},-{1\over2},-{1\over3},-{2\over3},
    -{1\over6})&\cr\ilspace
&&v_3&&(-{1\over6},-{1\over2},0,0,-{1\over2})&\cr\ilspace
&&v_4&&(-{2\over9},0,-{1\over3},-{2\over3},-{2\over3})&\cr\ilspace
&&v_5&&(-{1\over9},0,-{2\over3},-{1\over3},-{1\over3})&\cr\ilspace
\tablerule\noalign{\vskip2pt}
\multispan{5}\hfil {\tenpoint Table 3a: Generators for first cone}\hfil\cr
}
}\hskip5mm
\def\ilspace{\omit&height1.5pt&&&&\cr}
\vbox{\tabskip=0pt \offinterlineskip
\halign{\strut#&
\vrule#&\hfil\quad#\quad\hfil&\vrule#&\hfil\quad#\quad\hfil&
\vrule#\cr \tablerule\ilspace
&&Resolution&&Generators&\cr\ilspace
\tablerule\ilspace
&&1&&$v_1$, $v_2$, $v_3$, $v_4$, $v_5$&\cr\ilspace
&&2&&$v_1$, $v_1-v_2+v_3+v_4$, $v_3$, $v_4$, $v_5$&\cr\ilspace
&&3&&$v_1$, $v_2$, $v_2-v_3+v_4$, $v_4$, $v_5$&\cr\ilspace
&&4&&$v_1$, $v_2$, $v_3$, $v_2+v_3-v_4+v_5$, $v_5$&\cr\ilspace
&&5&&$v_1$, $v_2$, $v_3$, $v_2+v_3-v_4+v_5$, &\cr
\omit&&&&
\hphantom{$v_1$, $v_2$, }
$v_3+(v_2+v_3-v_4+v_5)-v_5$&\cr\ilspace
\tablerule\noalign{\vskip2pt}
\multispan{5}\hfil \tenpoint Table 3b: Generators for all cones\hfil\cr
}}$$
}
\iffigs
$$\vbox{\centerline{\epsfxsize=7cm\epsfbox{mult-f3.ps}}
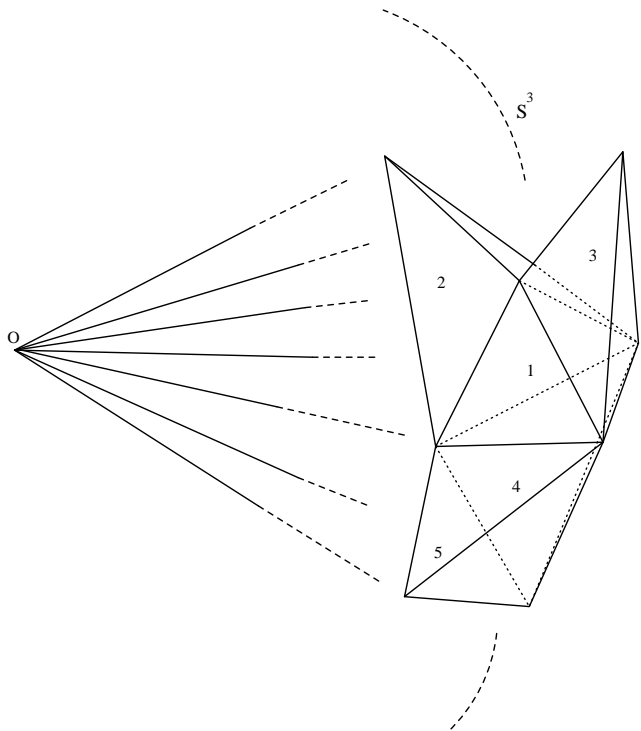
\centerline{Figure 3. Cones in the secondary fan.}}$$
\fi

\subsec{Predictions and their verification}

In table 4 we give the values of the invariant ratios of $(2,1)$
Yukawa couplings on $\X/G$ for each of the five large complex structure
choices singled out above.
We calculate the ratios as functions of $s$, where the coefficients
$a_i$ take the form $a_i=s^{r_i}$ for a vector $\vec{r}$ of exponents.
The vector of exponents $\vec{r}=\log|\vec{a}|=-\log|\vec{w}|$ must
lie in the specified cone after a sign change; we found such vectors
by simply adding the 5 generators of each cone (and then changing the
overall sign).

We evaluated the resulting function numerically for several large integer
values of $s$, and thereby determined the leading order terms as $s$
goes to infinity, displayed in table 4.
(We suppress the calculation in cases in which the topological
term in the expansion is 0/0, and leave the corresponding entries
in the table blank, since in those cases the leading order term
in $s$ measures an instanton contribution rather than a topological term.)
We see that the values of these ratios spectacularly
agree with the corresponding intersection forms in table 1
with the subleading term in $s$ corresponding to instanton corrections
on  $\M$. This is
extremely convincing evidence that each possible resolution is
physically realized.
Furthermore, not only is each possible resolution physically
realized, but as discussed in section 2
 we can continuously move from one such large complex structure
limit to any other {\it without encountering a physical singularity\/}
even though one necessarily encounters a classical geometric singularity
in the mirror description resulting in topology change.

{\eightpoint\hfuzz=10cm
\def\doots{\hbox to 4truept{$\dots$}}
\def\lquad{\hskip.3em\relax}
\def\tablerule{\noalign{\hrule}}
\def\ilspace{\omit&height2pt&&&&&&&&&&&&\cr}
$$\vbox{\tabskip=0pt \offinterlineskip
\halign{\strut#&
\vrule#&\hfil\quad$#$\quad\hfil&\vrule#&\hfil\lquad$#$\quad\hfil&
\vrule#&\hfil\lquad$#$\quad\hfil&\vrule#&\hfil\lquad$#$\quad\hfil&
\vrule#&\hfil\lquad$#$\quad\hfil&\vrule#&\hfil\lquad$#$\quad\hfil&
\vrule#\cr \tablerule\ilspace
&&\hbox{Resolution}&&1&&2&&3&&4&&5&\cr\ilspace\tablerule\ilspace
&&\hbox{Exponents}&&\omit\hfil$
({11\over9},1,{4\over3},{5\over3},{5\over3})
$\hfil&&\omit\hfil$
({7\over6},{1\over2},1,1,{5\over2})
$\hfil&&\omit\hfil$
({3\over2},{1\over2},2,3,{3\over2})
$\hfil&&\omit\hfil$
({13\over9},2,{5\over3},{4\over3},{4\over3})
$\hfil&&\omit\hfil$
({11\over6},{7\over2},1,1,{3\over2})
$\hfil&\cr\ilspace\tablerule\ilspace
&&{\langle \varphi_1^3\rangle\langle \varphi_4^3\rangle\over\langle
\varphi_1^2\varphi_4\rangle\langle \varphi_1\varphi_4^2\rangle}
&& -7-181s^{-1}+\doots &&  &&  && -{2\over5}s^2-{129\over250}s+\doots
&& 9+289s^{-1}+\doots &\cr\ilspace
&&{\langle \varphi_2^2\varphi_4\rangle\langle \varphi_3^2\varphi_4
\rangle\over\langle \varphi_2\varphi_3\varphi_4\rangle\langle
\varphi_2\varphi_3\varphi_4\rangle}
&& 2-5s^{-1}+\doots && 4-22s^{-1}+\doots && 0+2s^{-1}+\doots &&  &&
&\cr\ilspace
&&{\langle \varphi_2\varphi_3\varphi_4\rangle\langle \varphi_0\varphi_2^2
\rangle\over\langle \varphi_2^2\varphi_4\rangle\langle \varphi_0\varphi_2
\varphi_3\rangle}
&& 1+{1\over2}s^{-1}+\doots && 1+{3\over2}s^{-2}+\doots
&& 1+4s^{-1}+\doots && 0-2s^{-1}+\doots &&  &\cr\ilspace
&&{\langle \varphi_2\varphi_3\varphi_4\rangle\langle \varphi_0\varphi_1^2
\rangle\over\langle \varphi_1^2\varphi_4\rangle\langle \varphi_0\varphi_2
\varphi_3\rangle}
&& 2+27s^{-1}+\doots && 1-{1\over2}s^{-1}+\doots && -2s-33+\doots &&
&& 0+4s^{-2}+\doots &\cr
\tablerule
\noalign{\vskip2pt}
&\multispan{13}\hfil \tenpoint
Table 4: Asymptotic ratios of 3-point functions\hfil\cr
}}$$
}

\newsec{Conclusions}

The classical algebraic geometry of Calabi-Yau manifolds finds direct
expression in the lowest order approximations to the corresponding
$N = 2$ superconformal nonlinear sigma models. Numerous (but not all) aspects
of the latter physical systems receive quantum mechanical contributions
which cause them to differ from their classical mathematical values
giving rise to what might be called  ``quantum algebraic geometry''.
In this letter we have elucidated an important property of quantum
algebraic geometry: a singular operation in classical algebraic geometry
can have a quantum description which is perfectly smooth. We have seen
this in the particular context of topology changing processes.
Mirror symmetry has played a crucial role in this study as it allows us
to rephrase intractable questions on one space into fully understood
questions on its mirror.

In the present study we have seen that a set of topologically distinct
Calabi-Yau spaces smoothly fit together as conformal field theories.
In fact, the mirror description (which can be exactly described using
classical algebraic geometry) involves no topology change at all. We therefore
see that topology changing processes can be realized simply by
appropriately changing the expectation values of certain
truly marginal operators.

We have
focused in this letter on a particularly tractable and enlightening example
in which there are five topologically distinct Calabi-Yau spaces involved.
A less tractable, but more familiar example is the quintic hypersurface in
$\CP4$. In this case, there are {\it hundreds\/} of different resolutions
of the mirror quintic. As one  probes the  moduli space
of complex structures on the quintic, therefore, one is simultaneously
probing hundreds of topologically distinct Calabi-Yau spaces in the
mirror description. Thus, the full conformal field theory moduli space
requires not two but hundreds of topologically distinct Calabi-Yau
spaces for its complete geometric interpretation.

In the present letter we have emphasized
the large radius part of
the moduli space merely because the suppression of instanton effects
makes our calculations easier. One might wonder what exactly happens to
correlation functions as we pass through a wall (which, since a rational
curve is being blown down, necessarily has significant instanton corrections).
{}From the complex structure description on the mirror,  correlation
functions are manifestly
perfectly continuous. In the K\"ahler description, we know
leading order terms (the intersection form) can change. Thus, for our
picture to be correct, it must be that instanton corrections also change
upon passing through a wall in an {\it exactly compensating manner}.
A simple calculation, which
arose in conversations with E.~Witten \rnewW,
confirms
that this is in fact the case---as it must to be consistent with
the work presented here.
The physical system of the string---and the corresponding quantum algebraic
geometry of
the underlying Calabi-Yau sigma model---is thus completely well behaved
through a topology changing transition.

\bigbreak\bigskip\bigskip\centerline{{\bf Acknowledgements}}\nobreak
We thank R. Plesser for important
contributions at the inception and early stages of
this work.  We gratefully acknowledge numerous invaluable discussions
we have had with E.~Witten regarding both the present study and
his closely related work \rnewW.
We also thank S.~Katz, S.-S.~Roan, and R.~Wentworth for discussions.
The work of P.S.A.\ was supported by DOE grant
DE-FG02-90ER40542, the work of B.R.G.\ was supported
by the Ambrose Monell Foundation and by a National Young Investigator award,
and the work of D.R.M.\ was supported  by NSF grant DMS-9103827
and by an American Mathematical Society Centennial Fellowship.

\listrefs

\bye